\RequirePackage{lineno}
\documentclass[aps,prl,showpacs,superscriptaddress,twocolumn]{revtex4}  
\usepackage{graphicx}  
\usepackage{dcolumn}   

\usepackage{bm}        
\usepackage{ulem}      
\usepackage{color}     %
\usepackage{amssymb}   
\usepackage{multirow}
\usepackage{epstopdf}  

\newcommand {\etvis}        {\mbox{$E_T^{\rm vis}$}}
\newcommand {\gev}          {\mbox{$\rm GeV$}}
\newcommand {\gevc}         {\mbox{$\rm GeV\!/{\it c}$}}
\newcommand {\gevcsq}       {\mbox{$\rm GeV\!/{\it c}^2$}}
\newcommand {\mll}          {\mbox{$M_{\rm ll}$}}
\newcommand {\mllmin}       {\mbox{$M^{\rm min}_{\rm ll}$}}
\newcommand {\msnutau}      {\mbox{$M_{\snutau}$}}
\newcommand {\mvistau}      {\mbox{$M^{\rm vis}_{\tau}$}}
\newcommand {\pt}           {\mbox{$p_T$}}
\newcommand {\ppbar}        {\mbox{$p\bar{p}$}}
\newcommand {\pvistau}      {\mbox{$p^{\rm vis}_{\tau}$}}
\newcommand {\rparity}      {\mbox{$R_{p}$}}

\newcommand {\snutau}       {\mbox{$\widetilde{\nu}_{\tau}$}}

\begin{document}

\setpagewiselinenumbers \modulolinenumbers[1] 

\hspace{5.2in}  
\mbox{\hfill FERMILAB-PUB-10-077-E}

\title{Search for R-parity Violating Decays of $\tau$ Sneutrinos to $e\mu$, $\mu\tau$, and $e\tau$ Pairs in
\ppbar\ Collisions at $\sqrt{s} = 1.96$ TeV}

 \affiliation{Institute of Physics, Academia Sinica, Taipei, Taiwan 11529, Republic of China} 
\affiliation{Argonne National Laboratory, Argonne, Illinois 60439, USA} 
\affiliation{University of Athens, 157 71 Athens, Greece} 
\affiliation{Institut de Fisica d'Altes Energies, Universitat Autonoma de Barcelona, E-08193, Bellaterra (Barcelona), Spain} 
\affiliation{Baylor University, Waco, Texas 76798, USA} 
\affiliation{Istituto Nazionale di Fisica Nucleare Bologna, $^{dd}$University of Bologna, I-40127 Bologna, Italy} 
\affiliation{Brandeis University, Waltham, Massachusetts 02254, USA} 
\affiliation{University of California, Davis, Davis, California 95616, USA} 
\affiliation{University of California, Los Angeles, Los Angeles, California 90024, USA} 
\affiliation{University of California, San Diego, La Jolla, California 92093, USA} 
\affiliation{University of California, Santa Barbara, Santa Barbara, California 93106, USA} 
\affiliation{Instituto de Fisica de Cantabria, CSIC-University of Cantabria, 39005 Santander, Spain} 
\affiliation{Carnegie Mellon University, Pittsburgh, Pennsylvania 15213, USA} 
\affiliation{Enrico Fermi Institute, University of Chicago, Chicago, Illinois 60637, USA}
\affiliation{Comenius University, 842 48 Bratislava, Slovakia; Institute of Experimental Physics, 040 01 Kosice, Slovakia} 
\affiliation{Joint Institute for Nuclear Research, RU-141980 Dubna, Russia} 
\affiliation{Duke University, Durham, North Carolina 27708, USA} 
\affiliation{Fermi National Accelerator Laboratory, Batavia, Illinois 60510, USA} 
\affiliation{University of Florida, Gainesville, Florida 32611, USA} 
\affiliation{Laboratori Nazionali di Frascati, Istituto Nazionale di Fisica Nucleare, I-00044 Frascati, Italy} 
\affiliation{University of Geneva, CH-1211 Geneva 4, Switzerland} 
\affiliation{Glasgow University, Glasgow G12 8QQ, United Kingdom} 
\affiliation{Harvard University, Cambridge, Massachusetts 02138, USA} 
\affiliation{Division of High Energy Physics, Department of Physics, University of Helsinki and Helsinki Institute of Physics, FIN-00014, Helsinki, Finland} 
\affiliation{University of Illinois, Urbana, Illinois 61801, USA} 
\affiliation{The Johns Hopkins University, Baltimore, Maryland 21218, USA} 
\affiliation{Institut f\"{u}r Experimentelle Kernphysik, Karlsruhe Institute of Technology, D-76131 Karlsruhe, Germany} 
\affiliation{Center for High Energy Physics: Kyungpook National University, Daegu 702-701, Korea; Seoul National University, Seoul 151-742, Korea; Sungkyunkwan University, Suwon 440-746, Korea; Korea Institute of Science and Technology Information, Daejeon 305-806, Korea; Chonnam National University, Gwangju 500-757, Korea; Chonbuk National University, Jeonju 561-756, Korea} 
\affiliation{Ernest Orlando Lawrence Berkeley National Laboratory, Berkeley, California 94720, USA} 
\affiliation{University of Liverpool, Liverpool L69 7ZE, United Kingdom} 
\affiliation{University College London, London WC1E 6BT, United Kingdom} 
\affiliation{Centro de Investigaciones Energeticas Medioambientales y Tecnologicas, E-28040 Madrid, Spain} 
\affiliation{Massachusetts Institute of Technology, Cambridge, Massachusetts 02139, USA} 
\affiliation{Institute of Particle Physics: McGill University, Montr\'{e}al, Qu\'{e}bec, Canada H3A~2T8; Simon Fraser University, Burnaby, British Columbia, Canada V5A~1S6; University of Toronto, Toronto, Ontario, Canada M5S~1A7; and TRIUMF, Vancouver, British Columbia, Canada V6T~2A3} 
\affiliation{University of Michigan, Ann Arbor, Michigan 48109, USA} 
\affiliation{Michigan State University, East Lansing, Michigan 48824, USA}
\affiliation{Institution for Theoretical and Experimental Physics, ITEP, Moscow 117259, Russia} 
\affiliation{University of New Mexico, Albuquerque, New Mexico 87131, USA} 
\affiliation{Northwestern University, Evanston, Illinois 60208, USA} 
\affiliation{The Ohio State University, Columbus, Ohio 43210, USA} 
\affiliation{Okayama University, Okayama 700-8530, Japan} 
\affiliation{Osaka City University, Osaka 588, Japan} 
\affiliation{University of Oxford, Oxford OX1 3RH, United Kingdom} 
\affiliation{Istituto Nazionale di Fisica Nucleare, Sezione di Padova-Trento, $^{ee}$University of Padova, I-35131 Padova, Italy} 
\affiliation{LPNHE, Universite Pierre et Marie Curie/IN2P3-CNRS, UMR7585, Paris, F-75252 France} 
\affiliation{University of Pennsylvania, Philadelphia, Pennsylvania 19104, USA}
\affiliation{Istituto Nazionale di Fisica Nucleare Pisa, $^{ff}$University of Pisa, $^{gg}$University of Siena and $^{hh}$Scuola Normale Superiore, I-56127 Pisa, Italy} 
\affiliation{University of Pittsburgh, Pittsburgh, Pennsylvania 15260, USA} 
\affiliation{Purdue University, West Lafayette, Indiana 47907, USA} 
\affiliation{University of Rochester, Rochester, New York 14627, USA} 
\affiliation{The Rockefeller University, New York, New York 10021, USA} 
\affiliation{Istituto Nazionale di Fisica Nucleare, Sezione di Roma 1, $^{ii}$Sapienza Universit\`{a} di Roma, I-00185 Roma, Italy} 

\affiliation{Rutgers University, Piscataway, New Jersey 08855, USA} 
\affiliation{Texas A\&M University, College Station, Texas 77843, USA} 
\affiliation{Istituto Nazionale di Fisica Nucleare Trieste/Udine, I-34100 Trieste, $^{jj}$University of Trieste/Udine, I-33100 Udine, Italy} 
\affiliation{University of Tsukuba, Tsukuba, Ibaraki 305, Japan} 
\affiliation{Tufts University, Medford, Massachusetts 02155, USA} 
\affiliation{Waseda University, Tokyo 169, Japan} 
\affiliation{Wayne State University, Detroit, Michigan 48201, USA} 
\affiliation{University of Wisconsin, Madison, Wisconsin 53706, USA} 
\affiliation{Yale University, New Haven, Connecticut 06520, USA} 
\author{T.~Aaltonen}
\affiliation{Division of High Energy Physics, Department of Physics, University of Helsinki and Helsinki Institute of Physics, FIN-00014, Helsinki, Finland}
\author{J.~Adelman}
\affiliation{Enrico Fermi Institute, University of Chicago, Chicago, Illinois 60637, USA}
\author{B.~\'{A}lvarez~Gonz\'{a}lez$^w$}
\affiliation{Instituto de Fisica de Cantabria, CSIC-University of Cantabria, 39005 Santander, Spain}
\author{S.~Amerio$^{ee}$}
\affiliation{Istituto Nazionale di Fisica Nucleare, Sezione di Padova-Trento, $^{ee}$University of Padova, I-35131 Padova, Italy} 

\author{D.~Amidei}
\affiliation{University of Michigan, Ann Arbor, Michigan 48109, USA}
\author{A.~Anastassov}
\affiliation{Northwestern University, Evanston, Illinois 60208, USA}
\author{A.~Annovi}
\affiliation{Laboratori Nazionali di Frascati, Istituto Nazionale di Fisica Nucleare, I-00044 Frascati, Italy}
\author{J.~Antos}
\affiliation{Comenius University, 842 48 Bratislava, Slovakia; Institute of Experimental Physics, 040 01 Kosice, Slovakia}
\author{G.~Apollinari}
\affiliation{Fermi National Accelerator Laboratory, Batavia, Illinois 60510, USA}
\author{J.~Appel}
\affiliation{Fermi National Accelerator Laboratory, Batavia, Illinois 60510, USA}
% ...

\collaboration{CDF Collaboration\footnote{With visitors from $^a$University of Massachusetts Amherst, Amherst, Massachusetts 01003,
$^b$Universiteit Antwerpen, B-2610 Antwerp, Belgium, 
$^c$University of Bristol, Bristol BS8 1TL, United Kingdom,
$^d$Chinese Academy of Sciences, Beijing 100864, China, 
$^e$Istituto Nazionale di Fisica Nucleare, Sezione di Cagliari, 09042 Monserrato (Cagliari), Italy,
$^f$University of California Irvine, Irvine, CA  92697, 
$^g$University of California Santa Cruz, Santa Cruz, CA  95064, 
$^h$Cornell University, Ithaca, NY  14853, 
$^i$University of Cyprus, Nicosia CY-1678, Cyprus, 
$^j$University College Dublin, Dublin 4, Ireland,
$^k$University of Edinburgh, Edinburgh EH9 3JZ, United Kingdom, 
$^l$University of Fukui, Fukui City, Fukui Prefecture, Japan 910-0017,
$^m$Kinki University, Higashi-Osaka City, Japan 577-8502,
$^n$Universidad Iberoamericana, Mexico D.F., Mexico,
$^o$University of Iowa, Iowa City, IA  52242,
$^p$Kansas State University, Manhattan, KS 66506,
$^q$Queen Mary, University of London, London, E1 4NS, England,
$^r$University of Manchester, Manchester M13 9PL, England,
$^s$Muons, Inc., Batavia, IL 60510, 
$^t$Nagasaki Institute of Applied Science, Nagasaki, Japan, 
$^u$University of Notre Dame, Notre Dame, IN 46556,
$^v$Obninsk State University, Obninsk, Russia,
$^w$University de Oviedo, E-33007 Oviedo, Spain, 
$^x$Texas Tech University, Lubbock, TX  79609, 
$^y$IFIC(CSIC-Universitat de Valencia), 56071 Valencia, Spain,
$^z$Universidad Tecnica Federico Santa Maria, 110v Valparaiso, Chile,
$^{aa}$University of Virginia, Charlottesville, VA  22906,
$^{bb}$Bergische Universit\"at Wuppertal, 42097 Wuppertal, Germany,
$^{cc}$Yarmouk University, Irbid 211-63, Jordan,
$^{kk}$On leave from J.~Stefan Institute, Ljubljana, Slovenia, 
}}
\noaffiliation

\begin{abstract}

  We present a search for $\tau$ sneutrino production using the Tevatron $\ppbar$
collision data collected with the CDF II detector and corresponding to an
integrated luminosity of 1~${\rm fb^{-1}}$. We focus on the scenarios predicted by the
R-parity violating (RPV) supersymmetric models in which $\tau$ sneutrinos decay
to two charged leptons of different flavor. With the data consistent with
the standard model expectations, we set the upper limits on $\sigma(\ppbar \rightarrow \snutau) \times \rm BR (\snutau \rightarrow e\mu,\mu\tau,e\tau)$ and use these results to constrain the RPV couplings as a function
of $\tau$ sneutrino mass.
\end{abstract}

\maketitle
Supersymmetric (SUSY) extensions of the standard model (SM) are 
among the leading candidates for a theory which can solve the hierarchy problem and 
provide a framework for unifying particle interactions~\cite{Martin:1997ns}. 
Gauge-invariant and renormalizable interactions introduced in 
the SUSY models can violate the conservation of baryon ($B$) and lepton ($L$) number 
and lead to a proton lifetime shorter than the current experimental 
limits~\cite{proton_decay}. 
This problem is usually solved by postulating conservation 
of an additional quantum number, R-parity $\rparity=(-1)^{3(B-L)+2s}$, 
where $s$ is the particle spin~\cite{Farrar:1978}. 
However, models with R-parity-violating (RPV) interactions conserving spin and 
either $B$ or $L$ can also avoid direct contradiction with the 
proton lifetime upper limits~\cite{proton_lifetime_pdg}. 
Such models have the advantage that they naturally introduce
lepton flavor violation and can generate non-zero neutrino masses and 
angles~\cite{M_A_Diaz:2005} consistent with neutrino-oscillation 
data~\cite{Ashie:2004mr}. They can also explain the recently reported 
anomalous phase of the $b$ to $s$ transition~\cite{A.Kundu:2008}. 
From an experimental standpoint, RPV interactions allow for single 
production of supersymmetric particles (sparticles) in high-energy particle collisions and for
sparticles to decay directly into SM particles only; this makes the lightest 
sparticle unstable and critically affects the experimental strategy 
of the SUSY searches. 
Due to their clean final-state signatures, processes of single slepton 
production followed by decay to a pair of SM charged leptons become 
promising search channels for R-parity violating SUSY particles~\cite{Dimopoulos:1998}.

In this Letter we report a search for a heavy $\tau$ sneutrino \snutau\,  produced in quark-antiquark $d\bar{d}$ annihilation and decaying via lepton-flavor-violating interactions into $e\mu / \mu\tau / e\tau$ final states. 
The search is performed
using data corresponding to an integrated luminosity of 1~$\rm{fb}^{-1}$  collected in \ppbar\ collisions at
$\sqrt{s} = 1.96$ TeV by the CDF II detector at the Tevatron. The results 
are analysed in the framework of the minimal supersymmetric extensions 
of the SM~\cite{Martin:1997ns}, where the RPV part of the superpotential relevant to the $\tau$ sneutrino production and decay can be written as 
\begin{eqnarray} 
W_{RPV}& = & \lambda^{'ijk}
\textbf{L}_i \textbf{Q}_j \bar{\textbf{d}}_k 
 +  \frac{1}{2} \lambda^{ijk}
\textbf{L}_i \textbf{L}_j \bar{\textbf{e}}_k
\label{eq:W_Rp_odd} 
\end{eqnarray} 
$\textbf{L}$ and $\textbf{Q}$ in Eq.(\ref{eq:W_Rp_odd}) 
are the $SU(2)$ doublet superfields of leptons and quarks; 
$\bar{\textbf{e}}$, $\bar{\textbf{u}}$, and $\bar{\textbf{d}}$ 
are the $SU(2)$ singlet superfields of leptons, u-type and d-type quarks; 
$\lambda'$ and $\lambda$
are the Yukawa couplings at the production and decay vertex respectively; 
the indices $i$, $j$, and $k$ denote the fermion generations. We assume  single-coupling dominance
and the third super-generation to be the lightest.
The couplings $\lambda'_{311}= 0.10$ and $\lambda_{i3k} = 0.05$, compatible with the current indirect limits~\cite{Barbier:2004ez}, 
are chosen as a benchmark point. Heavy sneutrinos have been extensively searched for at LEP~\cite{Barbier:2004ez}. Recently, searches for heavy sneutrinos decaying into the $e \mu$
final state have been performed by the
CDF~\cite{Abulencia:2006xm} and D0
collaborations~\cite{Abazov:2007zz}.  
The results in this paper supersede~\cite{Abulencia:2006xm}.
This analysis also represents the first search for lepton-flavor-violating 
decays of heavy sneutrinos into final states involving a third 
generation lepton, the $\tau$, at the Tevatron.

CDF II is a general-purpose particle detector, described 
in detail elsewhere~\cite{cdfdet}. This measurement uses information 
from the central tracker~\cite{cot}, calorimeters~\cite{cdf_cem_ces,cdf_cha}, 
and muon detectors~\cite{cdf_muon_system} for charged lepton 
reconstruction and identification. 
Reconstruction of photons and $\pi^0$ mesons makes 
extensive use of the CES, the central shower maximum detector 
which is embedded at a depth of six radiation lengths within the 
electromagnetic calorimeter~\cite{cdf_cem_ces}. 
The luminosity is measured by a hodoscopic system of Cherenkov counters~\cite{clc}.
The event geometry and kinematics are described using the azimuthal 
angle $\phi$ around the beamline and the pseudorapidity 
$\eta = -\ln \tan \frac{\theta}{2}$, where $\theta$ is the polar angle 
with respect to the beamline. The transverse energy and momentum of 
the reconstructed particles 
are defined as: 
$ E_{T} = E \sin\theta$, $ \pt = p \sin\theta$, 
where $E$ is the energy and $p$ is the momentum. 

The data used in this measurement are collected using inclusive
high-\pt\ electron and muon triggers which select high-\pt\ electron and muon candidates with ${\rm |\eta|} \lesssim 1.0$.  
After event reconstruction, electron and muon candidates with $\pt \geq ~20~\gevc$ are identified using the procedures described in~\cite{elemuoid}. 
In addition we use independent measurements of the electron energy in the CES
to improve the overall electron selection efficiency and identification of electron candidates radiating significant energy due to the bremsstrahlung. 
The $\tau$ leptons are identified via their hadronic decays 
as narrow calorimeter clusters associated with one or three charged 
tracks~\cite{cdf_ztautau_prd_2007}. 
As the neutrino from the $\tau$ decay escapes detection, the ``visible'' 
four-momentum of a $\tau$ candidate, \pvistau, is reconstructed summing 
 the four-momenta of charged particle tracks and neutral particle
CES showers with a pion mass hypotheses. 
The resolution in \pvistau\ is further improved 
by combining measurements of the track momenta and energies of the CES showers with the 
energy measurements in the calorimeter. A reconstructed $\tau$ candidate is required 
to have the visible transverse energy, 
$\etvis$, greater than $ 25 ~\gev$ and its most energetic track must have 
$\rm \pt > 10 ~\gevc$. 
The invariant mass of its decay products, $\mvistau ~=~ \sqrt{\pvistau^2}$, is required 
to be consistent with the $\tau$ lepton decay: $\rm \mvistau ~<~ (1.8 + 0.0455\times(\etvis/GeV - 20))~\gevcsq$, where 
the second term in the formula accounts for a degradation of the resolution 
in \mvistau\ at high energy. Reconstructed $\tau$ candidates within $\Delta R = \sqrt{{\Delta \eta}^{2}+{\Delta \phi}^{2}}< 0.4$
of an identified electron or of a loosely-identified muon with large electromagnetic energy deposit in the calorimeter
are excluded from the analysis.

Events selected for the analysis are required to have two identified 
central ($|\eta| < 1$) lepton candidates of different flavor and opposite 
electric charge. The leptons have to be isolated: the extra energy 
measured within a cone of radius $\Delta R \leq 0.4$ surrounding the leptons must be
less than 10\% of the lepton energy. Events with leptons consistent 
with a photon conversion or a cosmic ray hypothesis are removed from the
analysis sample~\cite{elemuoid}. 

Signal and background studies are performed using Monte Carlo (MC) samples 
generated by {\sc pythia}~\cite{pythia} with the Tune A of CTEQ5L 
parton distribution functions~\cite{cteq}. 
The detector response is simulated with a {\sc geant3}-based package~\cite{geant}.  
The trigger, reconstruction and identification efficiencies are measured using $Z$ events
as calibration samples~\cite{elemuoid}. 

The predicted yield of signal events is calculated
using the next-to-leading order (NLO)  $\ppbar\rightarrow\snutau$ production cross section~\cite{ShaoMing:2006dx}. The total $\snutau$ width is defined by the $d\bar{d}$ and $l_il_k$ decay modes as
$\Gamma_{\snutau} = (3 {\lambda'}_{311}^2 + 2 \lambda_{i3k}^2 )M_{\snutau}/{16\pi}$,  where \msnutau is \snutau\ mass. 


There are several sources of background events that pass our analysis selections. We classify these
contributions based on whether the lepton candidates reconstructed in these events originated from a
``real" lepton (produced from a W or Z decay) or were a result of a hadron being misidentified as a
lepton, lepton flavor mis-assignment or a secondary lepton inside a jet. We collectively
refer to the lepton candidates of the second category as ``fakesÕ" and classify each contributing background
process into Type I, II and III according to the typical number of real and fake leptons reconstructed. Type I
contains events with two real leptons and includes $Z/\gamma^{*}\rightarrow\tau\tau$, 
diboson ($WW$, $WZ$, $ZZ$) and $t\bar{t}$ events. Type I is therefore called physics background. Type II includes events with one 
reconstructed fake lepton. They come from either (i) the $W/Z/ \gamma^{*}+$jet(s) events where one of the reconstructed leptons is in fact a jet misidentified as a lepton or (ii) $Z/\gamma^{*}\rightarrow ee/\mu\mu$ events with one of the leptons misidentified as a
lepton of a different flavor. The backgrounds in Type I and II are estimated using MC, and their expected event 
yields are normalized to the NLO cross sections~\cite{drell-yan,diboson,toppair}.
Events with two fake leptons (Type III) are dominated by multi-jet events with two jets 
misidentified as leptons and $\gamma +$jets events; in the latter case, a converted 
photon is not identified as such and gets reconstructed as an electron and a jet is misidentified 
as a $\mu$ or a $\tau$. The contribution of the processes in Type III is estimated 
using a data sample with two leptons of the same-charge and assuming 
no charge correlation between the two misidentified leptons.  Both Type II and III are called fake background.  

The systematic uncertainties in this search arise from a number of sources. 
The uncertainty on the luminosity 
measurement is 6\%~\cite{luminosity}. Uncertainties on lepton
identification efficiency are 3\% for $\tau$'s, 1\% for electrons, and 1\% 
for muons~\cite{elemuoid}. The jet-to-$\tau$ misidentification probability is measured with 
an accuracy of 15\%. Uncertainties in the parton distribution functions (PDF) result in 
 the systematic error on the predicted signal cross section, which 
 varies from 4\% to 20\% and increases with the $\snutau$ mass. Variations 
 of the signal acceptance due to PDF uncertainties are less than 1\%.
 
\begin{figure}[htbp]
\begin{center}
\includegraphics[width=0.35\textwidth]{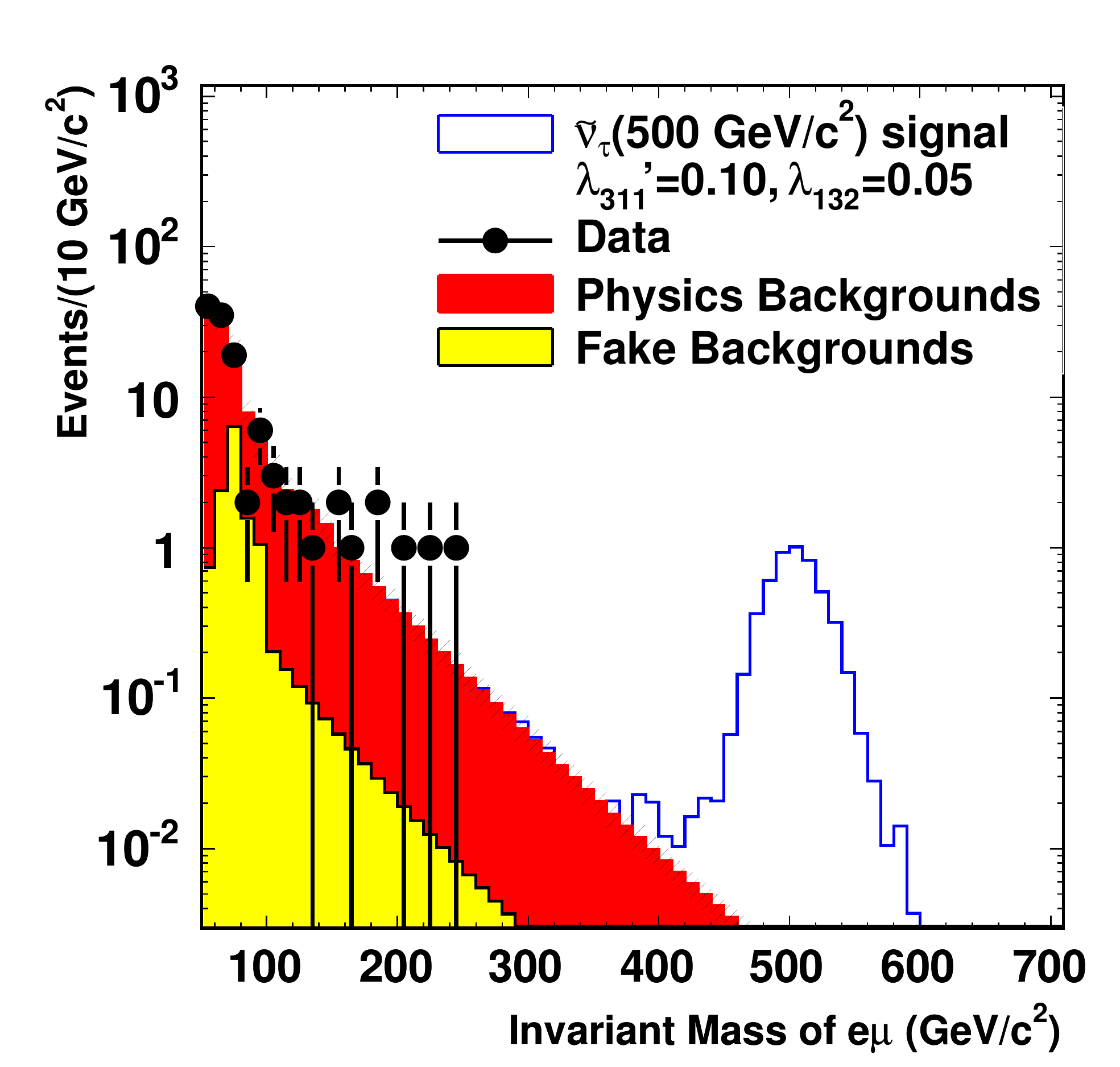}
\includegraphics[width=0.35\textwidth]{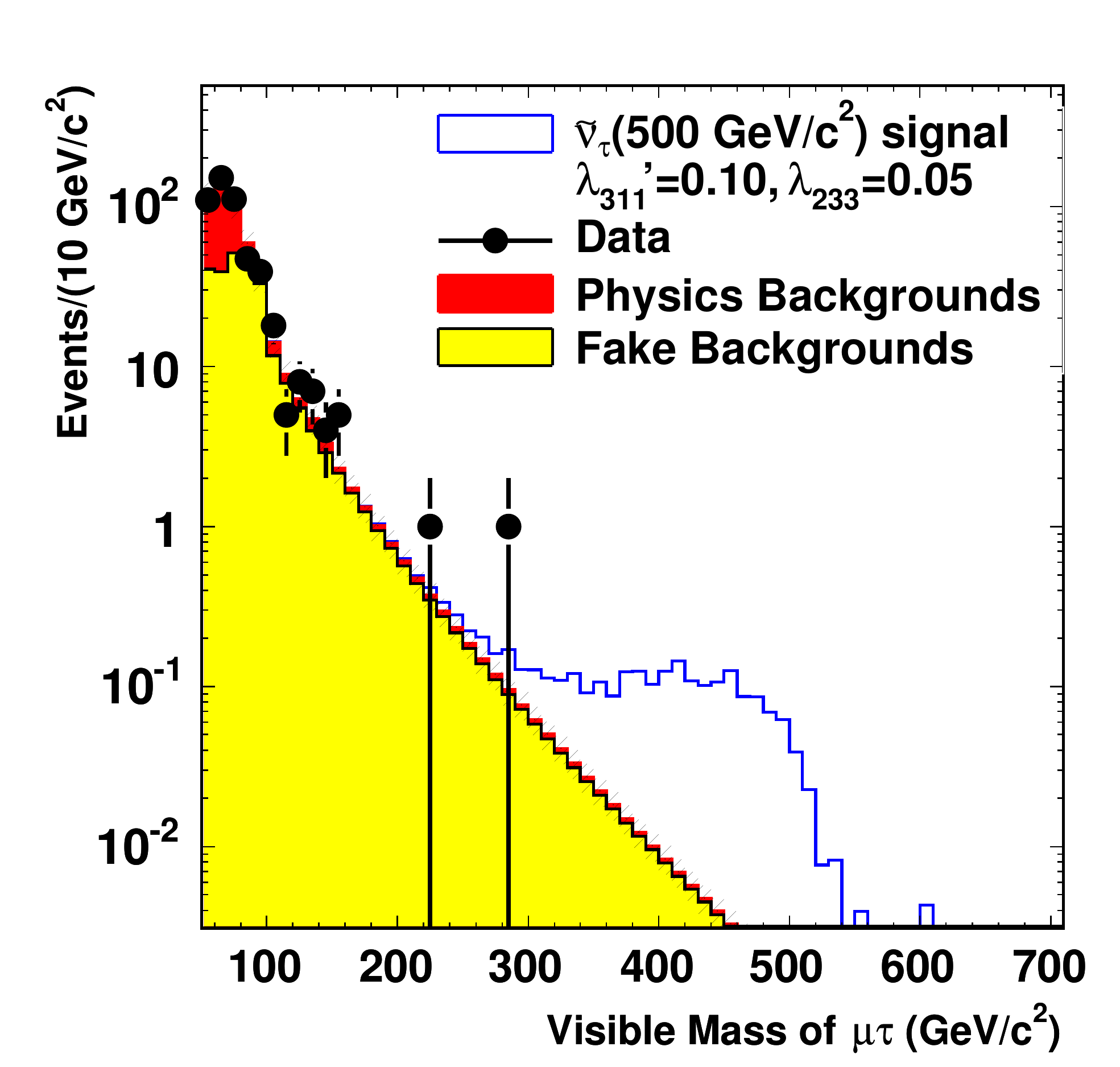}
\includegraphics[width=0.35\textwidth]{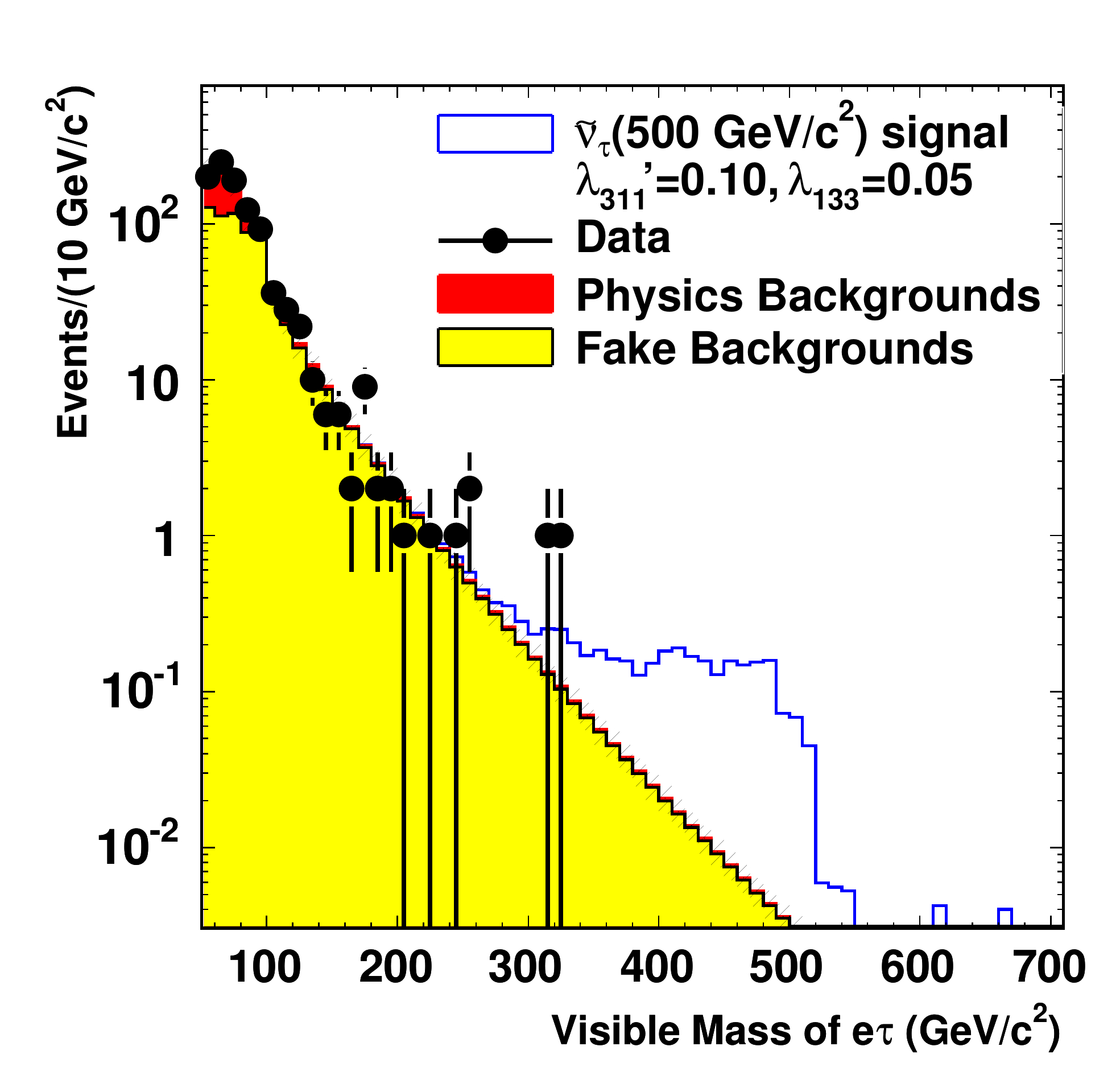}

\end{center}

\caption[] {
  Expected SM and observed distributions in \mll\ 
  for $e\mu$, $\mu\tau$ and $e\tau$ channels. 
  Also shown is an expected \snutau\ signal for \msnutau = 500 \gevcsq and  RPV couplings $\lambda'_{311}= 0.10$ and $\lambda_{i3k} = 0.05$. 
  \label{fig:observation}
}
\end{figure}

We search for a signal from $\snutau$ decays into lepton pairs in the
distributions for dilepton invariant mass, $\mll$ (note that in the case
of a hadronically decaying $\tau$, it is the $\tau$ visible energy that is used to
calculate the mass). The low mass region,
$50~\gevcsq < \mll < 110~\gevcsq$, is used to validate the event selection
and the background normalization. The observed and expected event
yields in this region are in good agreement, as summarized in Table~\ref{tab:controlregion}.
With the normalization fixed, the backgrounds are extrapolated 
into the region $\mll > 100~\gevcsq$, where the search is performed as
a ``blind" counting experiment. 
Figure~\ref{fig:observation} compares data distributions in $\mll$ to the SM expectations 
for each of the three channels. With no statistically significant 
excesses observed, we use the data to set 
upper limits on 
$\sigma (\ppbar\rightarrow\snutau)\times {\rm BR}(\snutau \rightarrow e\mu/\mu\tau/e\tau)$.
\begin{table}[htbp]
  \centering{  
    \caption{ 
      The observed and predicted event yields in the control region. Uncertainties
      on the predicted yields include both statistical and systematic contributions.
      \label{tab:controlregion}
    }

    \vspace{1mm} 
\begin{tabular}{cccccccc}
\hline
\hline \multicolumn{4}{c}{Control Region    \ \ \ \ 
\ \ \  $\rm 50 ~\gevcsq < \mll < 110 ~\gevcsq$}
\\
 Channel                   & e$\mu$                       & $\mu \tau$         & e$\tau$ \\
 \hline
 Physics Backgrounds     & \multirow{1}{*}{100.2$\pm$6.5} &\multirow{1}{*}{262.4$\pm$21.0} &\multirow{1}{*}{309.6$\pm$24.7}                \\
 Fake Backgrounds     & \multirow{1}{*}{9.4$\pm$3.3}& \multirow{1}{*}{222.6$\pm$31.7} & \multirow{1}{*}{577.8$\pm$37.0}                \\

Total SM                                       & 109.6$\pm$7.7 & 485.0$\pm$40.9   & 887.4$\pm51.0$\\
 Observed                           & 105    & 477                                               & 894            \\
\hline
\hline
\end{tabular}
}
\end{table}



\begin{figure}[htbp]
\hspace{-0.4in}
\includegraphics[width=0.5\textwidth]{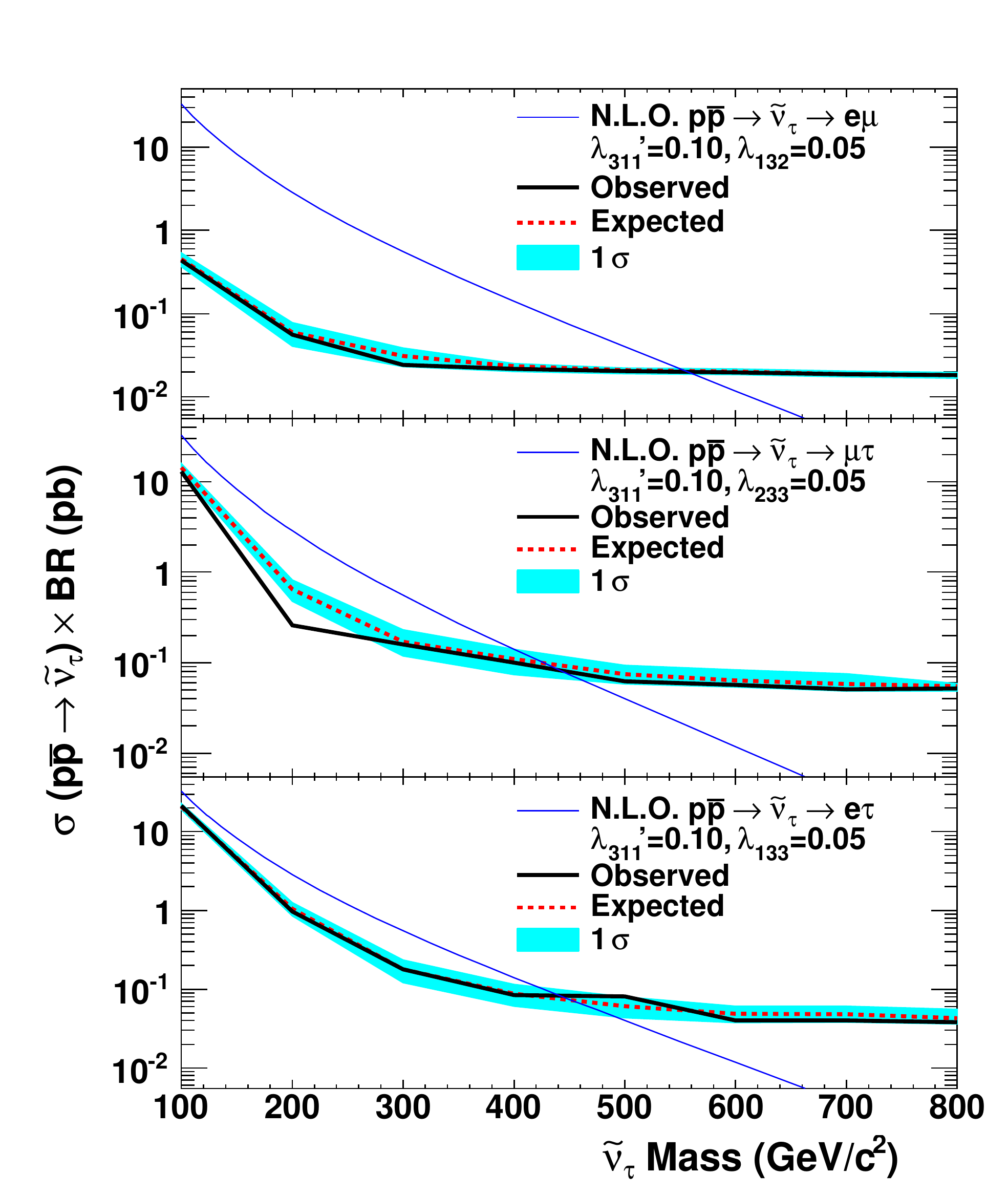}

\caption[]
{
  The expected and observed 95\% C.L. upper limits on $\sigma (\ppbar\rightarrow\snutau)\times {\rm BR}(\snutau \rightarrow e\mu/\mu\tau/e\tau)$ 
  as a function of \msnutau.
  \label{fig:expsens}
}
\end{figure}

In each of the channels ($e \mu$, $\mu\tau$, $e\tau$), the expected
and observed upper limits 
are calculated using a Bayesian technique~\cite{joel} at 95\% credibility 
level (C.L.) as a function of \msnutau. 
For a given \msnutau , the limits are calculated 
by integrating the differential cross section $d\sigma/d\mll$ over the 
region $\mll > \mllmin$, where \mllmin\ optimizes the search 
sensitivity for a selected \msnutau.
The search results for \msnutau\ = 500~\gevcsq\ are summarized in 
Table ~\ref{table:sigbgmutau}.
Figure~\ref{fig:expsens} shows the expected and observed 95\% C.L. 
upper limits on $\sigma (\ppbar\rightarrow\snutau)\times {\rm BR}(\snutau \rightarrow e\mu/\mu\tau/e\tau)$ as a function of \msnutau.  
We also set 95\% C.L. upper limits on $ {\lambda'_{311}}^{2} \times {\rm BR}(\snutau \rightarrow e\mu/\mu\tau/e\tau)$ as shown in Table~\ref{table:limit3}. 
\begin{table}[htbp]
\centering{
  \caption{
    Expected and observed number of events in $e\mu$, $\mu\tau$, and $e\tau$ 
    channels. The expected yields of \snutau\ events are calculated for \msnutau = 500 \gevcsq 
    and RPV couplings $\lambda'_{311}= 0.10$ and $\lambda_{i3k} = 0.05$.
    \label{table:sigbgmutau}
  }

\begin{tabular}{cccccccc} 
\hline
\hline
Channel                                         &   e$\mu$          & $\mu \tau$      & e$\tau $       \\ 
$\mllmin (~\gevcsq)$                      & $  440 $           & $ 300 $           & $ 310 $       \\
\hline Physics Backgrounds     &        \multirow{1}{*}{0.03$\pm0.01$ }                                  &    \multirow{1}{*}{0.1$\pm$0.02 }                           &   \multirow{1}{*}{0.2$\pm$0.03 }              \\
 Fake Backgrounds     &   \multirow{1}{*}{0.01$\pm$0.01 }                                    &        \multirow{1}{*}{0.3$\pm$0.1 }                       &      \multirow{1}{*}{0.6$\pm$0.1 }    \\
                                                                                                              
Total SM background                  & 0.04 $\pm$ 0.01   & 0.4$\pm$ 0.1    & 0.9$\pm$0.1 \\
Expected signal                            & 5.9 $\pm$ 0.1       & 2.0$\pm$0.1       & 2.7$\pm$0.1   \\ 
Observed                                       &       0                        &      0                      &      2        \\ \hline \hline
\end{tabular}
}
\end{table}

\begin{table}[htbp]
\centering{
  \caption{95\% C.L. upper limits on $ {\lambda'_{311}}^{2} \times {\rm BR}(\snutau \rightarrow e\mu/\mu\tau/e\tau)$.     
  \label{table:limit3}
  }
\begin{tabular}{cccccccc} 
\hline
\hline
$\msnutau( ~\gevcsq)$                      &  $e\mu $          &   $ \mu \tau $     &    $ e\tau $    \\
\hline    300                  &     $ \ 6 \times 10^{-5} \   $     & $ \ 4 \times 10^{-4} \ \ $   & $\ 5 \times 10^{-4}  \ $ \\
              400                  &     $ \ 2 \times 10^{-4}  \ $    & $\ 1  \times 10^{-3}   \ $ & $\  9\times 10^{-4} \ $\\
              500                  &    $ \ 7 \times 10^{-4} \ $     &  $ \ 2\times 10^{-3}  \ $  & $ \ 3\times 10^{-3} \ $\\
              600                  &    $ 2 \times 10^{-3}$    &  $ 7\times 10^{-3} $ &  $5\times 10^{-3}$ \\
              700                  &     $8\times 10^{-3} $    & $ 2\times 10^{-2} $   &  $2\times 10^{-2} $\\
              \hline \hline
\end{tabular}
}
\end{table}
In conclusion, we have searched for production of a massive sneutrino 
decaying to $e\mu$, $\mu\tau$, or $e\tau$ final states via R-parity violating 
interactions. We find the data consistent with the SM predictions and calculate 
the 95\% C.L. upper limits on the $\sigma (\ppbar\rightarrow\snutau)\times {\rm BR}(\snutau \rightarrow e\mu/\mu\tau/e\tau)$ in the mass range 
up to 800 \gevcsq. Using these cross section limits, we constrain $ {\lambda'_{311}}^{2} \times {\rm BR}(\snutau \rightarrow e\mu/\mu\tau/e\tau)$ as a function of \msnutau. This analysis sets to the first Tevatron limits for lepton-flavor violating  decays of heavy sneutrinos into final states
involving a third generation lepton. For the RPV couplings $\lambda'_{311}= 0.10$ and $\lambda_{i3k} = 0.05$ 
the observed 95\% C.L. lower limits on \snutau\ mass are 
558 \gevcsq\ in the $e \mu$ channel,
441 \gevcsq\ in the $\mu \tau$ channel, and 
442 \gevcsq\ in the $e \tau$ channel. 

We thank the Fermilab staff and the technical staffs of the participating institutions for their vital contributions.
%
%
This work was supported by the U.S. Department of Energy and National Science Foundation; the Italian Istituto Nazionale di Fisica Nucleare; the Ministry of Education, Culture, Sports, Science and Technology of Japan; the Natural Sciences and Engineering Research Council of Canada; the National Science Council of the Republic of China; the Swiss National Science Foundation; the A.P. Sloan Foundation; the Bundesministerium f\"ur Bildung und Forschung, Germany; the World Class University Program, the National Research Foundation of Korea; the Science and Technology Facilities Council and the Royal Society, UK; the Institut National de Physique Nucleaire et Physique des Particules/CNRS; the Russian Foundation for Basic Research; the Ministerio de Ciencia e Innovaci\'{o}n, and Programa Consolider-Ingenio 2010, Spain; the Slovak R\&D Agency; and the Academy of Finland.




\end{document}